\documentclass[aps,prl,twocolumn,showpacs]{revtex4}

\newcommand{\Jeff}{J_{\mbox{\footnotesize eff}}}
\newcommand{\re}{\mbox{\cal Re}}

\usepackage{graphicx}

\begin{document}

\title{Instability and control of a periodically-driven
Bose-Einstein condensate}

\author{C.E.~Creffield}
\affiliation{Dpto de F\'isica de Materiales, Universidad
Complutense de Madrid, E-28040, Madrid, Spain}

\date{\today}

\begin{abstract}
We investigate the dynamics of a Bose-Einstein condensate
held in an optical lattice under the influence of a strong periodic
driving potential. Studying the mean-field version of the Bose-Hubbard 
model reveals that the condensate becomes highly unstable
when the effective intersite tunneling becomes negative.
We further show how controlling the sign of the tunneling can be
used as a powerful tool to manage the dispersion of an atomic wavepacket, 
and thus to create a pulsed atomic soliton laser.
\end{abstract}


\maketitle

{\em Introduction -- }
The spectacular experimental progress in confining Bose-Einstein condensates
in optical lattice potentials has provided a powerful tool for
investigating many-body quantum dynamics. Such optical potentials
are extremely clean and controllable, and together with
their long decoherence times, this allows the observation of many
coherent lattice phenomena
which are extremely challenging to study in other solid-state systems.
One such effect is ``coherent destruction of tunneling'' (CDT)
\cite{hanggi}, observed very recently
in atomic systems \cite{morsch, oberthaler},
in which a periodic driving field acts to renormalise
the tunneling between lattice sites. This control
over the dynamics of the condensate is achieved without 
altering any of the parameters of the optical lattice,
and has natural
applications to quantum information, since it preserves
the system's coherence. However, 
it is crucial to know the stability of 
the condensate during its time evolution, particularly
the presence of {\em dynamical instability}, in which
deviations from a steady state grow exponentially with time.
The case of a static potential was analyzed in Ref.\cite{wu},
and studied experimentally in Refs.\cite{burger,instability_expt},
and it was found that dynamical instability occurs above a certain
critical quasimomentum. Later work \cite{accel} examined the case of
a uniformly accelerated lattice, and found the non-intuitive result
that dynamical instability was enhanced in the limit of low
acceleration. 

In this paper we build upon this approach to analyze the richer
and more complex case 
where the optical potential is {\em periodically} rocked.
To achieve this we carry out the stability
analysis about the Floquet states \cite{hanggi} of the system, 
which are the appropriate
generalization of energy eigenstates to the case of a time-periodic
Hamiltonian. An important point for experiment is to minimize
or avoid instabilities, and so we first find the critical
interaction strength at which dynamical instability occurs. 
We then connect this with the behavior of the effective
tunneling $\Jeff$, and show how manipulating $\Jeff$ via CDT
can be used to control the dynamics of the condensate, providing
control over matter-wave dispersion \cite{dispersion} and thus
allowing the creation of bright solitons.

{\em Method -- }
A system of cold bosons held in an optical lattice can be
described very accurately \cite{jaksch} by the Bose-Hubbard
Hamiltonian
\begin{equation}
H_{\mbox{\footnotesize BH}} = -J \sum_{\langle m, n \rangle} 
\left( a_m^{\dagger} a_n + H.c. \right) + 
\frac{U}{2} \sum_m n_m \left( n_m - 1 \right) \ ,
\label{hubbard}
\end{equation}
where $a_m^{\dagger}$/$a_m$ are the boson creation/annihilation
operators, and $n_m = a_m^{\dagger} a_m$ is the number operator.
The properties of the system are governed by the
hopping parameter $J$, and the Hubbard interaction $U$ which
describes the potential energy between two bosons occupying
the same lattice site. An extremely valuable means of studying
and controlling such systems is to accelerate the
lattice by varying the phase-difference
between the two laser beams forming the standing wave potential.
In the rest frame of the lattice this acceleration manifests itself
as an inertial force which effectively ``tilts'' the potential.
If instead of a uniform acceleration the lattice is
periodically accelerated and decelerated, it is possible to
produce a potential that oscillates periodically in time,  
$H_I = K \cos \omega t \sum_m m n_m$. where
$K$ and $\omega$ parametrise its amplitude and frequency respectively.

In order to study the stability of the driven condensate we will
first pass to a mean-field description of
this model, analogous to the Gross-Pitaevskii equation, and
then linearize about the ground-state
to obtain the Bogoliubov equations for the condensate excitations.
We first write the Heisenberg equations
of motion for the boson operators $a_n$, and then 
take the classical field approximation and treat
them simply as c-numbers $\alpha_n$. It is
then straightforward to show that the classical amplitudes 
obey the equation of motion
\begin{equation}
i \frac{\partial \alpha_n}{\partial t} = -J 
\left( \alpha_{n+1} + \alpha_{n-1} \right) + g \left| \alpha_n \right|^2
+ K \cos \omega t \ n \alpha_n
\label{mean}
\end{equation}
where for convenience we have scaled the interaction as $g = U/N$. 
Note that we also take $\hbar=1$, and will measure all energies
in units of $J$.

To simplify the analysis we use periodic boundary conditions.
In the limit of large 
lattice sizes, however, the choice of boundary conditions does not affect 
the underlying physics, and we will later use Dirichlet boundary
conditions to simulate the time-evolution of the condensate.
In the absence of interactions, the eigenstates of the system will
simply be plane waves $\alpha_n = \exp\left[i n p\right]$ where
$p$ is the angular momentum. With this in mind, we take as a trial 
solution $\alpha_n = \exp\left[i \left( n \phi + \theta \right) \right]$,
where $\phi$ and $\theta$ are functions to be determined.
Substituting this solution in Eq.\ref{mean} yields the result
\begin{eqnarray}
\phi(t) &=& p - \frac{K}{\omega} \sin \omega t \\ 
\theta(t) &=& 2 J \left( \cos p \ S(t) - \sin p \ C(t) \right) - g t
\end{eqnarray}
where the functions $S(t) / C(t)$ are defined in terms of Bessel functions
as
\begin{eqnarray}
C(t) &=& \sum_{m = -\infty}^{\infty} \frac{\cos m \omega t - 1}{m \omega}
{\cal J}_m(K/\omega) , \\
S(t) &=& \sum_{m = -\infty}^{\infty} \frac{\sin m \omega t}{m \omega}
{\cal J}_m(K/\omega) .
\end{eqnarray}
 
\begin{center}
\begin{figure}
\includegraphics[width=0.45\textwidth,clip=true]{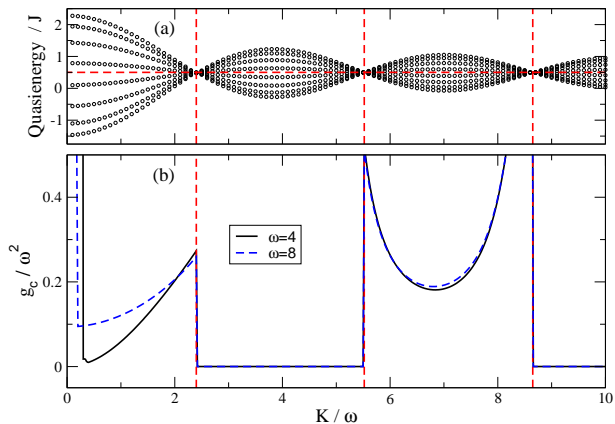}
\caption{(a) Quasienergy spectrum of the driven mean-field Bose-Hubbard
model, for an 8-site system with $\omega=16$. The width
of the spectrum is modulated by the Bessel function ${\cal J}_0(K/\omega)$
and displaced by the interaction energy $g=0.5$, in full agreement with the
analytical solution Eq.\ref{spectrum}.
(b) Plot of the critical interaction, $g_c$, at which
the system becomes dynamically unstable, for driving
frequencies $\omega=4$ and $8$. This quantity diverges at $K/\omega=0$
and at the Bessel function zeros (vertical dashed lines), and is
zero when $\Jeff$ is negative, for which dynamical instability occurs
for any positive value of $g$.}  
\label{stability}
\end{figure}
\end{center}

Since the Hamiltonian of the system is periodic in time, the Floquet
theorem dictates that the solutions of the time-dependent
Schr\'odinger equation can be written in the form
$\exp\left[i \epsilon t \right] u(t)$, where $\epsilon$ is
termed the quasienergy, and $u(t)$ is a $T$-periodic function
called the Floquet state. 
To obtain the quasienergies we thus simply have to
extract the terms from the solution which are not $T$-periodic,
giving the result
\begin{equation}
\epsilon(p) = 2 \cos p \ J {\cal J}_0\left(K/\omega\right) + g .
\label{spectrum}
\end{equation}
In the absence of the driving the quasienergies thus form a
normal single-particle bandstructure, the interaction $g$ acting merely 
to shift the entire spectrum. The driving then acts
to renormalise the width of the spectrum by the Bessel function
${\cal J}_0$, as was previously observed in a theoretical analysis 
\cite{holthaus} of semiconductor superlattice systems.
In Fig.\ref{stability}a we show numerical results
for the quasienergies of an 8-site system, obtained directly
from the time-evolution of the system, which beautifully
corroborate the expected behavior.
In particular, when the Bessel function becomes zero,
the spectrum collapses to a point and the system 
will manifest CDT. 

In order to analyze the dynamical stability of the ground state ($p=0$),
we now introduce a perturbation $\alpha_n(t)=\alpha_n^0(t)
(1 + u(t) \exp\left[ i q n \right] + v^\ast(t) \exp \left[ -i q n \right])$,
where $\alpha_n^0(t)$ is the unperturbed solution, and $q$ and
$\omega$ are the momentum and energy of the excitation.
We then linearize Eq.\ref{mean} about this solution to
obtain the Bogoliubov de Gennes equations for 
$u(t)$ and $v(t)$
\begin{equation}
i \frac{d}{dt}\left(\! \!
\begin{array}{c}
u(t) \\
v(t)
\end{array}
\! \! \right) = 
{\cal L}(q,t)
\!\left(\! \!
\begin{array}{c}
u(t)  \\
v(t)
\end{array}
\! \!\right) ,
\label{bogo}
\end{equation}
where the elements of the matrix ${\cal L}(q,t)$ are given by
\begin{eqnarray}
{\cal L}_{11}(q,t) &=& 4 J \sin \left( q/2 \right) \sin \left(q/2 
- K/\omega \sin \omega t \right) + g , \nonumber \\
{\cal L}_{12}(q,t) &=& g = - {\cal L}_{21}(q,t) , \nonumber \\
{\cal L}_{22}(q,t) &=& -4 J \sin \left( q/2 \right) \sin \left(q/2
+ K/\omega \sin \omega t \right) - g . \nonumber
\end{eqnarray}
We can note that, similarly to the Hamiltonian, the operator
${\cal L}(q,t)$ is $T$-periodic.
Consequently we can also apply the
Floquet theorem to describe the time-evolution of the quasiparticle excitation
$(u,v)$. To find the corresponding Floquet states, we numerically evolve
Eq.\ref{bogo} over one period of driving, using the $2 \times 2$
identity matrix as the initial state. The result of this procedure
is the single-period propagator $U$. The 
eigenstates of $U$ are then the excitation Floquet states,
while its eigenvalues are related to the excitation quasienergies via
$\lambda_i = \exp \left[ -i T \epsilon_i \right]$.

The symmetries of $U$, combined with the normalization condition
obeyed by the quasiparticle excitation, $|u|^2 - |v|^2 = 1$, allow
the characteristic equation to be written in the
particularly simple form
\cite{accel}
\begin{equation}
\lambda_\pm = \re \left[ U_{11} \right] \pm
\sqrt{ \re \left[ U_{11} \right]^2 -1 } .
\end{equation}
The condition of dynamical stability can now be recast in terms
of the eigenvalues of ${\cal L}(q,t)$, in analogy with the
use of Lyapunov exponents in classical mechanics.
The solution is stable against
an excitation with momentum $q$ if there is no quasienergy with
a positive imaginary component, or equivalently, if 
$\re \left[ U_{11} \right] < 1$. This suggests a simple scheme
to map out the stability zones of the driven system. For a given
choice of $K$ and $\omega$ we select a value of $g$ and scan
over the range of $q$. If $\re \left[ U_{11} \right] \leq 1$ for
all values of $q$ we can declare that the system is stable for
these parameters, and that to induce instability we need to
increase $g$ to a higher value. In this way a standard bisection
scheme can be used to locate the instability boundary, $g_c$.

{\em Results -- }
We show the results of this procedure for two driving frequencies in
Fig.\ref{stability}b. We can first observe that for $K/\omega = 0$
it is possible to directly diagonalise ${\cal L}(q)$ to obtain
the result 
$\epsilon_\pm = \pm 2 \sqrt{2} \sin \left( q/2 \right)
\sqrt{2 J^2  \sin^2 \left( q/2 \right) + J g }$. 
As expected, this duplicates the familiar result for the
Bogoliubov excitations of an undriven, stationary condensate \cite{wu_njp}.
It is also clear from
this expression that dynamical instability will not occur if the
interaction is repulsive, since the eigenfrequencies will not become complex 
unless the product $\left( J \ g  \right) < 0$.
Accordingly, as $K/\omega \rightarrow 0$ we can see
from Fig.\ref{stability}b that the value of $g_c$ diverges.
For larger $K/\omega$ the value of $g_c$ then rapidly drops, passing
through a broad local minimum before again diverging as $K/\omega$
approaches 2.4048 -- the first zero of ${\cal J}_0$.
This corresponds to the onset of CDT;
as the effective tunneling is reduced, the
dynamics of the condensate is suppressed, and stability is regained.

Passing through the zero of ${\cal J}_0$, we can see that dynamical
stability is then abruptly lost. In this region the condensate becomes
dynamically unstable for {\em any} positive value of the interaction.
We can obtain some insight into this effect from Eq.\ref{spectrum},
by defining an effective tunneling, $\Jeff = J {\cal J}_0(K/\omega)$.
When $K/\omega$ is increased from 2.4048,
the Bessel function changes sign and $\Jeff$ becomes negative. 
The physical significance of this sign-change
has been observed previously in experiment \cite{morsch}, 
where it caused the momentum 
distribution function to be discretely shifted by $\pi$.
This occurs because the tunneling, as well as being renormalized in
amplitude, acquires a phase-factor of $\exp[i \pi]$ \cite{current}.
Accordingly, if we view the driving field as acting simply to renormalise
the tunneling, the product $\left( \Jeff \ g \right)$ now becomes negative and
dynamical instability can indeed occur. 
A similar effect would occur if instead $g$ were made negative
(for example, by using a Feshbach resonance), for which the condensate
would become attractive and thus unstable toward collapse.
We can note that in Ref.\cite{morsch} the condensate
was close to non-interacting, and thus avoided this instability.
When $\Jeff$ becomes zero at $K/\omega=5.52$
we can see that $g_c$ again diverges due to CDT.
The same pattern of behavior then repeats.
Fig.\ref{stability}b also shows that
$g_c$ scales quite accurately as 
$g_c \sim \omega^2$. Surprisingly, the zone of 
stability thus becomes wider at high driving frequencies, although the 
acceleration of the lattice is much larger. A similar feature was
seen in the analysis of an accelerated condensate \cite{accel},
which found an increasing propensity to dynamical instability in the
limit of low acceleration. 

We have so far used an ideal flat optical lattice potential.
In experiment, however, an additional
harmonic trap potential is usually present which can
substantially modify the dynamics of the system, and introduce new effects.
To investigate this, we now apply
an additional quadratic potential $V =  k r_i^2$, where
$r_i$ is measured from the center of the system.
We initialize 
the system in the ground state of the the mean-field Hamiltonian 
(\ref{mean}) in the presence of the trap, and then
displace the trapping potential 
by a distance of $25$ lattice spacings, thereby exciting the condensate
into motion.

\begin{center}
\begin{figure}
\includegraphics[width=0.22\textwidth,clip=true]{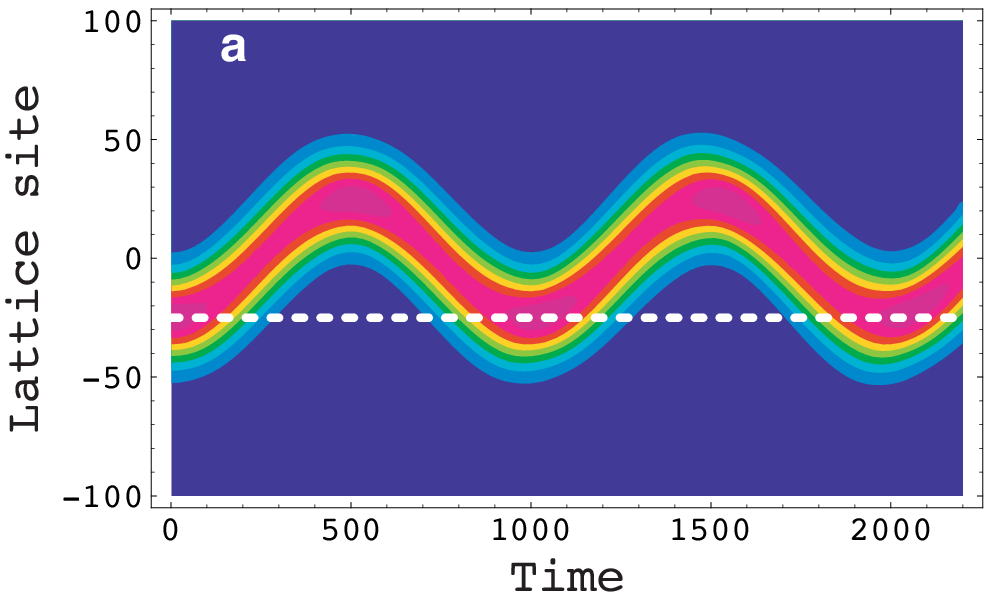}
\includegraphics[width=0.22\textwidth,clip=true]{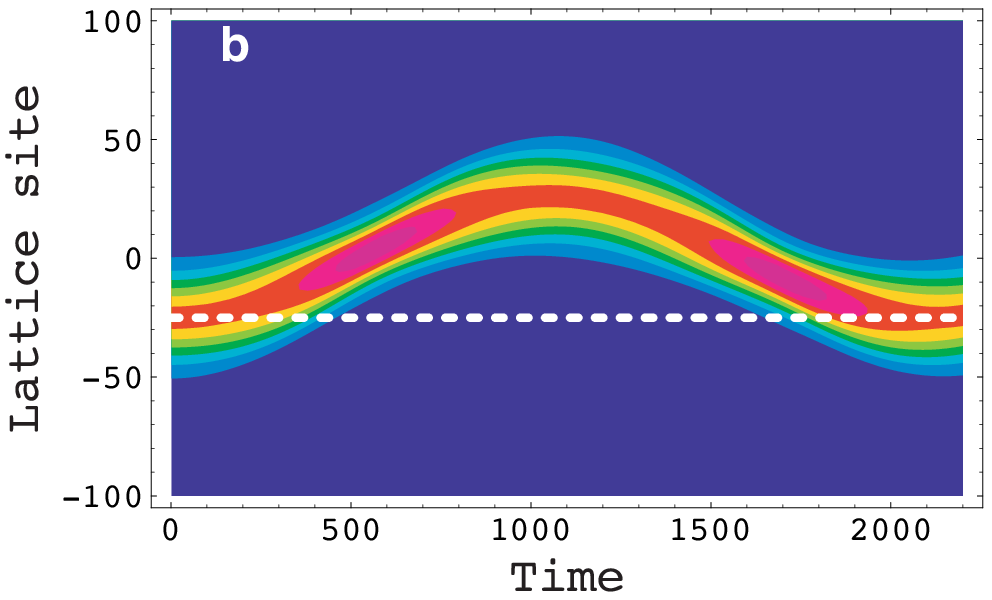}
\includegraphics[width=0.22\textwidth,clip=true]{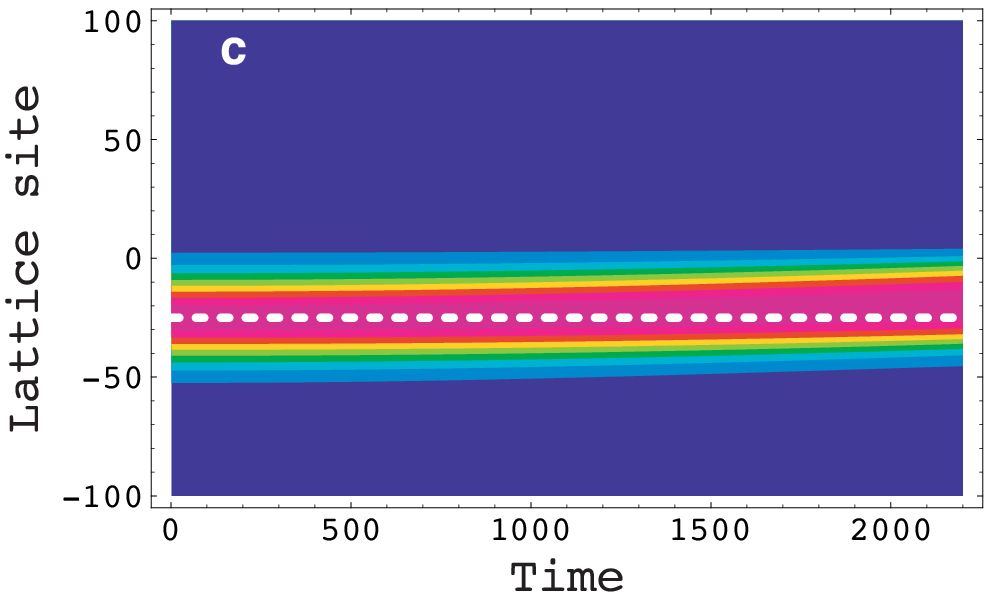}
\includegraphics[width=0.22\textwidth,clip=true]{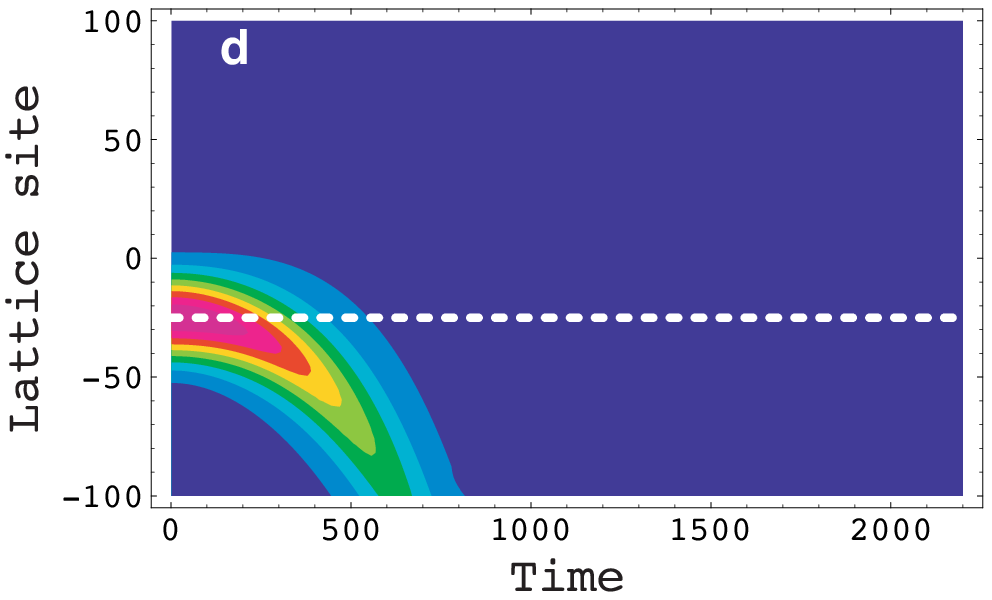}
\caption{Dynamics of a condensate in a parabolic trap. At $t=0$ the
trap is shifted by $25$ lattice spacings to induce the condensate
into motion. System parameters are $g=0.1$ and $\omega=4$. 
(a) In the absence of driving, the condensate sloshes periodically
from side to side of the trap with a well-defined frequency.
(b) For $K/\omega=2.2$, $\Jeff$ is reduced in amplitude, and
so the frequency of the oscillation is correspondingly reduced.
(c) For $K/\omega=2.40$, the tunneling is completely suppressed since 
$\Jeff \simeq 0$, and the time-evolution of the condensate is
thus frozen.
(d) For $K/\omega=3$, $\Jeff$ becomes negative. The trap potential thus
appears inverted to the condensate, which is rapidly expelled from the
center.}
\label{slosh}
\end{figure}
\end{center}

In Fig.\ref{slosh}a, we show the evolution of the condensate in the
absence of the periodic driving. The condensate makes
a periodic oscillation of constant amplitude, very similar to  
the center of mass motion observed in experiment \cite{burger}.
The period of oscillation
is governed by the intersite tunneling, or equivalently,
by the condensate's effective mass ($m^\ast \propto \Jeff^{-1}$), 
and thus allows these quantities to be measured directly. In Fig.\ref{slosh}b
the system is subjected to a driving with $K/\omega=2$. Clearly
the oscillation period has increased, corresponding to the
expected reduction of the tunneling by ${\cal J}_0$,
which can alternatively be interpreted as an enhancement of the
effective mass. Increasing $K/\omega$ further to the first zero of 
${\cal J}_0$ produces CDT, and 
so the system remains frozen in its initial state (Fig.\ref{slosh}c).
In this case the effective mass has become infinite.

A further increase of $K/\omega$ means that the Bessel function changes
sign, and thus $\Jeff$ becomes {\em negative}.
We have seen already that this sign-change has a significant effect
on the dynamical stability of the ground-state, and as
we show in Fig.\ref{slosh}d it has an equally dramatic effect on
the dynamics of the condensate.
Instead of oscillating, the condensate now rapidly accelerates away from 
the center of the trap. The reason for this becomes evident when we
examine the terms of the Hamiltonian. Making $\Jeff$ negative
is clearly equivalent to time-reversed evolution with a positive tunneling,
but with reversed signs for the trapping potential and non-linearity $g$.
Thus when $\Jeff$ changes sign, the condensate behaves as an
{\em attractive} condensate in an {\em inverted} potential, and so
is quickly expelled from the center of the trap.

Controlling the parameters of the periodic driving field thus allows
us to tune the effective mass of the condensate to be positive, negative,
or infinite. A particularly important application of this control
is the production of solitons. Solitons remain stable by balancing
dispersion with the interparticle interaction, and thus bright soliton
solutions of the Gross-Pitaevskii equation demand either using an
attractive interaction, or the use of complicated staggered 
phase-imprinting techniques. Making the effective mass negative,
however, would allow bright solitons to be created in repulsive
condensates, without having to reverse the sign of the interaction
or requiring phase imprinting.

\begin{center}
\begin{figure}
\includegraphics[width=0.45\textwidth,clip=true]{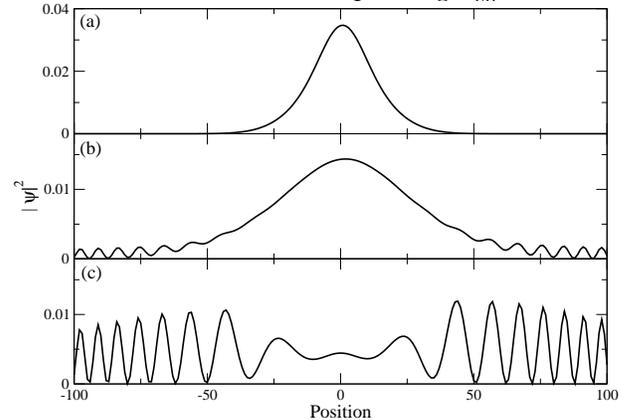}
\caption{Time-evolution of a condensate held in a parabolic trap,
subjected by a periodic driving field. System parameters are
$g=0.1$, and $\omega=4$. At $t=0$ the amplitude of the
tunneling is suddenly changed from $K/\omega=2.0$ to 3.80,
to reverse the sign of $\Jeff$.
(a) At $t=0$ the condensate has a gaussian density-profile.
(b) $t=200 T$, the condensate has spread, and has
developed oscillations at its edges.
(c) $t=400 T$ the oscillations have deepened into localized pulses,
similar to those seen in Ref.\cite{carr},
and the initial state has almost completely converted into
a soliton train.}
\label{soliton}
\end{figure}
\end{center}

To investigate this possibility, we revisit a proposal made by Carr
and Brand in Ref.\cite{carr} for manipulating a trapped condensate
to produce a train of solitons. The procedure consists of two parts;
the scattering length of the condensate is changed from positive
to negative while at the same time the trap is inverted to become
expulsive. As we have seen previously, both of these processes can be 
accomplished simultaneously by tuning the sign of $\Jeff$ from positive 
to negative. To verify this, the trapped system was again initialized in
its ground-state, and the amplitude of the driving slowly
increased from zero to $K/\omega=2.0$ so that the system adiabatically
followed to the ground-state of the renormalized Hamiltonian.
As can be seen in the top panel of Fig.\ref{soliton}, the resulting
density distribution is gaussian in form.
The amplitude of the driving was then abruptly altered to
$K/\omega=3.8$, reversing the sign of $\Jeff$.
The resulting time evolution of the condensate shown in Fig.\ref{soliton}
clearly displays the transformation of the condensate into a train of 
solitonic pulses, with the solitons forming first at the edges of the 
condensate.

{\em Conclusions -- } In summary, we have 
shown how an oscillating driving potential can be used
to renormalise the effective tunneling, $\Jeff$, of
a Bose-Einstein condensate. 
This parameter crucially determines when the condensate 
is dynamically unstable, and
when $\Jeff$ is negative the condensate becomes
unstable for any positive value of interaction. 
This is particularly important for the design of experiments,
since in such parameter ranges the interaction
must be set to an appropriately small value.
We have also shown how
manipulating $\Jeff$ in this way may also be used 
as a novel tool to to control
the dynamics of a condensate, in a complementary way
to the well-known method of controlling the interaction
via Feshbach resonances. This both extends the possibility
of manipulating the condensate to systems which do not possess
convenient resonances, and provides a new means
to investigate the interplay between
nonlinearity and dispersion, notably the production of solitons.

\smallskip
The author was supported by a Ram\'on y Cajal Fellowship.


\end{document}